\documentclass
[%
 reprint,
 amsmath,amssymb,
 aps,
]{revtex4-2}

\usepackage{graphicx}
\usepackage{dcolumn}
\usepackage{bm}
\usepackage{hyperref}
\usepackage{csquotes}
\usepackage{xcolor}

\usepackage{pgfplots}

\begin{document}

\title{Non-interacting holographic dark energy with Torsion via Hubble Radius}

\author{Yongjun Yun}
\affiliation{Graduate School Department of Physics, Daejin University, Pocheon 11159, Korea}

\author{Jungjai Lee}
\email{jjlee@daejin.ac.kr}
\affiliation{Graduate School Department of Physics, Daejin University, Pocheon 11159, Korea\\
School of Physics, Korea Institute for Advanced Study, Seoul 02455, Korea}

\begin{abstract}
We reconstruct a holographic dark energy model within a Friedmann cosmology incorporating torsion scalar, assuming no interaction between dark energy and dark matter.
Setting the Hubble radius as an infrared (IR) cut-off, we focus on a system dominated by contribution of a time-dependent torsion scalar induced by the spin of matter.
In this regime, our results show that even very weak torsion causes cosmic acceleration.
Specifically, we find that minima of the current equation of state for holographic dark energy, $(\omega_X^{0})_{min}$, lies in the range $-1 < (\omega_X^{0})_{min} < -0.778$ as a free parameter $d$ varies from $1$ to $0.654$.
Focusing on the free parameter $d \approx 1$, we find that $(\omega_X^{0})_{min}$ exhibits slightly different behavior from the cosmological constant.
Introducing torsion allows the Hubble radius to serve as a viable IR cut-off even without assuming the interaction between them. 
Moreover, this approach provides a non-interacting limit not found in earlier interacting models that use the Hubble radius as the IR cut-off.
\end{abstract}

\maketitle


\section{Introduction}
Since the discovery of the cosmic accelerating expansion \cite{1,2}, its source, dark energy, remains poorly understood, however it is interpreted as the cosmological constant in the $\Lambda$CDM model.
Recent data reveal the Hubble tension, a discrepancy between the expansion rates predicted by $\Lambda$CDM and measured through the local distance ladder \cite{3}.
This suggests the need for new dark energy models beyond $\Lambda$CDM, though none based on general relativity have proven to be problem-free \cite{4}.
Thus, exploring extended gravity theories, such as the Einstein-Cartan theory, which is the simplest classical extension of general relativity.

The Einstein-Cartan theory introduces torsion, a macroscopic manifestation of matter's intrinsic angular momentum (spin).
Given that elementary particles possess spin, its effects may be crucial to consider.
The classical limit of a quantum gravity theory should include the Einstein-Cartan theory, and the holographic principle should be embedded within it.
For example, string theory’s background $B$ field introduces torsion, while the metric field gives Riemann curvature, and the AdS/CFT correspondence \cite{5} is realized via holography.

Cohen et al. \cite{6} suggested that in a system of finite size $L$, vacuum energy should not exceed the mass of a black hole of the same size, and an infrared (IR) cut-off is related to an ultraviolet (UV) cut-off, which is known as UV/IR mixing or holographic principle. 
In the framework of general relativity, Li \cite{7} proposed the holographic dark energy model and explained the accelerating expansion of the universe by assuming that the IR cut-off relevant to dark energy is the future event horizon.
However, this approach has been criticized for issues related to causality and circular logic \cite{8}.
Alternatively, using the Hubble radius or the particle horizon as the IR cut-off 
fails to account for cosmic acceleration.
When an interaction between dark energy and dark matter is allowed, 
employing the Hubble radius as the IR cut-off can lead to acceleration.
Nevertheless, there exists no non-interacting limit \cite{9}.

In this Letter, we reconstruct the holographic dark energy model by incorporating torsion scalar, assuming no interaction between dark energy and dark matter.
Unlike previous work that interprets torsion as vacuum energy \cite{10}, we treat torsion as a purely geometric quantity, distinct from vacuum energy.
However, due to the very weak contribution of torsion in the current universe, the cosmological observations are still not sufficient to demonstrate an irrefutable discrepancy with the predictions of general relativity \cite{11}.
Given the lack of observational data on torsion, we examine three scenarios: (i) steady-state torsion, (ii) constant torsion scalar, and (iii) time-dependent torsion scalar linked to matter density with spin.
Interestingly, we show that introducing torsion allows the Hubble radius to be promoted as a candidate for the IR cut-off, even without interaction between dark energy and dark matter.
Consequently, cosmic acceleration can occur at the corresponding cut-off despite weak torsion, implying that issues with causality and circular logic can be avoided.

This paper is organized as follows: Section 2 introduces the Einstein-Cartan theory and the a cosmology with torsion scalar. Section 3 sets the Hubble radius to the IR cut-off and evaluates an equation of state for holographic dark energy in this cosmology. 

\section{Torsion scalar Cosmology}
In the Einstein-Cartan theory, the torsion tensor is geometrically defined as the antisymmetric part of the affine connection, $S^{\rho}{}_{\mu\nu} \equiv \Gamma^{\rho}{}_{[\mu\nu]}$, and it has the physical dimension of inverse length, i.e., $[S^{\rho}{}_{\mu\nu}] = [\mathrm{L}^{-1}]$.
Under the metricity condition, the connection satisfies the relation
$\Gamma^{\rho}{}_{\mu\nu} = \tilde{\Gamma}^{\rho}{}_{\mu\nu} + S^{\rho}{}_{\mu\nu} + 2S_{(\mu\nu)}{}^{\rho}$, where $\tilde{\Gamma}^{\rho}{}_{\mu\nu}$ is the Levi-Civita connection.
The total action is given by
\begin{equation} \label{action}
    S
    = \frac{c^{4}}{16 \pi G}\int d^{4}x\sqrt{-g}\,R(\Gamma) + \int d^{4}x\sqrt{-g}\mathcal{L}_{m},
\end{equation}
where $R(\Gamma)$ is the Ricci scalar constructed from the affine connection, and $\mathcal{L}_{m}$ is the Lagrangian density describing matter fields.
Varying the action \eqref{action} with respect to the metric tensor $g_{\mu\nu}$, one obtains the Einstein–Cartan equations, which can be rewritten in a mathematical form identical to that of the Einstein equations \cite{12}:
\begin{equation} \label{EC equations}
    R_{\mu\nu}(\Gamma) - \frac{1}{2}g_{\mu\nu}R(\Gamma)
    = \frac{8\pi G}{c^{4}} T_{\mu\nu}.
\end{equation}
In the expression of the equations \eqref{EC equations}, $R_{\mu\nu}(\Gamma)$ and $T_{\mu\nu}$ denote the Ricci–Cartan tensor and the canonical energy–momentum tensor, respectively, both of which are generally asymmetric.

Varying the action \eqref{action} with respect to a contorsion defined as $K^{\rho}{}_{\mu\nu} = S^{\rho}{}_{\mu\nu} + S_{\mu\nu}{}^{\rho} + S_{\nu\mu}{}^{\rho}$, one obtains the Cartan equations, which provide an algebraic relation connecting the torsion tensor to its source. Specifically, 
\begin{equation} \label{Cartan equations}
    S_{\mu\nu\rho} 
    = -  \frac{2\pi G}{c^{4}} \left(2s_{\nu\rho\mu}+g_{\rho\mu}s_{\nu}-g_{\mu\nu}s_{\rho}\right),
\end{equation}
where $s_{\nu\rho\mu}$ denotes a spin tensor of the matter, and $s_{\nu} \equiv s^{\alpha}{}_{\nu\alpha}$ is a corresponding spin vector.
In the absence of matter, spin does not exist, and thus torsion vanishes.
Hence, this work does not address scenarios in which matter is absent.

Let us introduce a form of the torsion tensor that preserves the cosmological principle \cite{13, 14}:
\begin{equation} \label{torsion scalar}
    S_{\mu\nu\rho}
    = \frac{2}{c^{2}}\phi h_{\mu[\nu}u_{\rho]},
\end{equation}
where $\phi = \phi(t)$ denotes a time-dependent scalar function, referred to as the torsion scalar, and has the physical dimension of inverse time, i.e., $[\phi] = [\mathrm{T}^{-1}]$. Here, $u^{\mu}$ is a timelike four-velocity vector, and $h_{\mu\nu} = g_{\mu\nu} + u_{\mu}u_{\nu}/c^{2}$ is a projection tensor orthogonal to $u^{\mu}$.
This relation shows that the torsion can be characterized by the torsion scalar. 
Using the Cartan equations \eqref{Cartan equations}, the spin tensor of the matter can be parameterized by the torsion scalar.
In the terms of the torsion tensor \eqref{torsion scalar}, we can employ the FLRW metric given by
\begin{equation} \label{FLRW metric}
    ds^{2} 
    = - c^{2}dt^{2} + a^{2}\left[\frac{dr^{2}}{1-
Kr^2} + r^{2}\left(d\theta^{2}+\sin^{2}\theta\varphi^{2}\right)\right],
\end{equation}
where $a = a(t)$ is the scale factor, and $K$ is the curvature parameter.
For convenience, we now use natural units where $c=\hbar=1$.
In the flat metric with $K = 0$, the Einstein-Cartan equations \eqref{EC equations} yield torsional analogues of the Friedmann equations \cite{13}:
\begin{equation} \label{Friedmann equation 1}
    \left(\frac{\dot{a}}{a}\right)^{2} + 4\phi^{2} + 4\left(\frac{\dot{a}}{a}\right)\phi
    = \frac{1}{3M_{p}^{2}}\rho,
\end{equation}
and
\begin{equation} \label{Friedmann equation 2}
    \frac{\ddot{a}}{a} + 2\dot{\phi} + 2\left(\frac{\dot{a}}{a}\right)\phi
    = - \frac{1}{6M_{p}^{2}}\left(\rho+3p\right),
\end{equation}
where $\rho$ and $p$ denote the total energy density and pressure, respectively, and $M_{p} \equiv 1 / \sqrt{8 \pi G}$ is the reduced Planck mass.
When $\phi = 0$, these equations reduce to the standard Friedmann equations.
In this work, we consider two energy species, vacuum energy and matter, which are labeled by $X$ and $m$, respectively, so that $\rho = \rho_{m} + \rho_{X}$ and $p = p_{m} + p_{X}$.
From the Friedmann equation \eqref{Friedmann equation 2}, the deceleration parameter, defined as $q\equiv-\ddot{a}a/\dot{a}^{2}$, takes the following form:
\begin{equation} \label{deceleration parameter}
    q
    = \frac{1}{6M_{p}^{2}H^{2}}\left(\rho+3p\right) + 2\frac{\dot{\phi}}{H^{2}} + 2\frac{\phi}{H},
\end{equation}
which shows that the torsion scalar can affect cosmic acceleration.
Using the Hubble parameter defined by $H = \dot{a}/a$, the Friedmann equation \eqref{Friedmann equation 1} can be rewritten as
\begin{equation} \label{Friedmann equation 1-1}
    H^{2}\left(1+2\frac{\phi}{H}\right)^{2}
    = \frac{1}{3M_{p}^{2}}\rho.
\end{equation}
Here, we define an effective critical density as follows:
\begin{equation} \label{critical density}
    \rho_{c}
    \equiv 3M_{p}^{2}H^{2}\left(1+2\frac{\phi}{H}\right)^{2}.
\end{equation}
With effective density parameters $\Omega_{m} = \rho_{m}/\rho_{c}$ and $\Omega_{X} = \rho_{X}/\rho_{c}$, the Friedmann equation \eqref{Friedmann equation 1} recasts
\begin{equation} \label{Friedmann equation 1 (density parameter)}
   \Omega_{m} + \Omega_{X} 
   = 1.
\end{equation}

Combining the two Friedmann equations \eqref{Friedmann equation 1} and \eqref{Friedmann equation 2}, we obtain the continuity equation (see Appendix A):
\begin{equation} \label{continuity equation}
    \dot{\rho} + 3H\left(\rho+p\right) + 2\phi\left(\rho+3p\right)
    = 0,
\end{equation}
which shows that the evolution of the total energy density is influenced by the torsion scalar.
When $\phi = 0$, this equation reduces to the standard continuity equation. 
Although some models allow for interactions between dark energy and dark matter during cosmic evolution, such interactions have not yet been clearly confirmed by observations.
Therefore, non-interacting models are preferred from a fundamental perspective, and we focus on them.
Accordingly, the continuity equation \eqref{continuity equation} can be split into
\begin{equation} \label{Non-interaction (vacuum)}
    \dot{\rho}_{X} + 3H\left( 1+\omega_{X})\rho_{X} + 2\phi(1+3\omega_{X} \right)\rho_{X} 
    = 0,
\end{equation}
and
\begin{equation} \label{Non-interaction (matter)}
    \dot{\rho}_{m} + 3H\left( 1+\omega_{m})\rho_{m} + 2\phi(1+3\omega_{m} \right)\rho_{m}
    = 0,
\end{equation}
where $\omega_{X} = p_{X}/\rho_{X}$ and $\omega_{m} = p_{m}/\rho_{m}$ denote
the equations of state for vacuum energy and matter, respectively.
Here, we consider vacuum energy without spin and matter with spin.

\section{Holographic dark energy model with Torsion scalar}
Cohen et al. suggested that in a system of finite size $L$, vacuum energy should not exceed the mass of a black hole of the same size, which leads to an inequality $L^{3}\Lambda^{4} \lesssim LM_{p}^{2}$.
Here, $L$ and $\Lambda$ are the IR cut-off and the UV cut-off, respectively. Since the quantum zero-point energy density is $\rho_{X} \sim \Lambda^{4}$, thus $L^{3}\rho_{X} \lesssim LM_{p}^{2}$.
For the largest $L$ allowed, saturating the inequality, we obtain the holographic constraint:
\begin{equation} \label{hologrpahic constraint}
    \rho_{X} 
    = 3d^{2}M_{p}^{2}L^{-2},
\end{equation}
where $\rho_{X}$ is regarded as dark energy density, and $d$ is a free parameter.
In this work, we set the IR cut-off to the Hubble radius $L = H^{-1}$, so the constraint becomes
\begin{equation} \label{holographic constraint_Hubble}
    \rho_{X} 
    = 3d^{2}M_{p}^{2}H^{2}.
\end{equation}
In this context, the Friedmann equation \eqref{Friedmann equation 1} yields the matter density that takes the form:
\begin{equation} \label{matter density}
    \rho_{m} 
    = 3M_{p}^{2}H^{2}\left[\left(1+2\frac{\phi}{H}\right)^{2}-d^{2}\right].
\end{equation}
When $\phi = 0$, we get $\rho_{m} = 3M_{p}^{2}H^{2}(1-d^{2})$, which exhibits physically meaningful scaling behavior for $d^2 < 1$, but becomes unphysical at $d^2 = 1$, imposing an upper bound on $d$.
Moreover, we find the relation $\rho_{m} \sim H^{2} \sim \rho_{X}$, which indicates that the dark energy behaves like matter, i.e., $\omega_{X} = \omega_{m}$, so no cosmic acceleration occurs. 
Because of this result, the Hubble radius has been ruled out as a viable IR cut-off candidate within the framework of general relativity. 
On the other hand, following \eqref{matter density}, one can show that the presence of the torsion scalar prevents $\rho_{X}$ from behaving like $\rho_{m}$ and allows $d^{2} = 1$.
Thus we can superficially infer that $\omega_{X} \neq \omega_{m}$, which suggests the possibility that cosmic acceleration occurs due to the effect of the torsion scalar.

To obtain a concrete form of the equation of state for dark energy, we rewrite its continuity equation \eqref{Non-interaction (vacuum)} as
\begin{equation} \label{continuity equation_vacuum 1}
   \frac{d\ln{\rho_{X}}}{d\ln{a}}
   = - 3\left(1+\omega_{X}\right) - 2\frac{\phi}{H}\left(1+3\omega_{X}\right).
\end{equation}
Inserting the relation $\rho_{X} = \Omega_{X}\rho_{c} = \Omega_{X}(\rho_{m}/\Omega_{m}) = \Omega_{X}\rho_{m}/(1-\Omega_{X})$ derived from the Friedman equation \eqref{Friedmann equation 1 (density parameter)} into the left-hand side of \eqref{continuity equation_vacuum 1}, we obtain the following form of
\begin{equation} \label{equation of state_vacuum}
    \omega_{X} 
    = \omega_{m} - \frac{1}{3\left(1+2\frac{\phi}{H}\right)}\frac{\Omega_{X}^{'}}{\Omega_{X}\left(1-\Omega_{X}\right)},
\end{equation}
where the prime symbol indicates the derivative with respect to $\ln a$. 
Here, we used the continuity equation of matter \eqref{Non-interaction (matter)}.
Note that the holographic constraint \eqref{holographic constraint_Hubble} has not yet been imposed in the expression of the equation of state \eqref{equation of state_vacuum}.
We have the density parameter of form:
\begin{equation} \label{effective density parameter_Hubble}
    \Omega_{X} 
    = \frac{\rho_{X}}{\rho_{c}}
    = \frac{3d^{2}M_{p}^{2}H^{2}}{3M_{p}^{2}H^{2}\left(1+2\frac{\phi}{H}\right)^{2}}
    = \frac{d^{2}}{\left(1+2\frac{\phi}{H}\right)^{2}}.
\end{equation}
Putting \eqref{effective density parameter_Hubble} into \eqref{equation of state_vacuum}, we obtain the equation of state as follows:
\begin{equation} \label{equation of state_Hubble}
    \omega_{X} 
    = \omega_{m} + \frac{4}{3}\frac{1}{\left(1+2\frac{\phi}{H}\right)^{2}-d^{2}} 
    \left[ \frac{\dot{\phi}}{H^{2}}+(1+q)\frac{\phi}{H} \right].
\end{equation}
In order to express the equation of state in terms of dimensionless parameters, we introduced the deceleration parameter $q = -1-(\dot{H}/H^{2})$.
This equation of state can be determined only when both $d$ and the explicit form of $\phi$ are known.
Determining the free parameters theoretically is very challenging, but they have been calculated as $d = 0.886$ \cite{15} and $d \simeq 0.95$ \cite{16} from the quantum perspective and inferred as $d = 1$ \cite{17} from the thermodynamic perspective.
Based on these findings, we will now discuss the case where the free parameter is positive, i.e., $d > 0$.
It should be noted that these are based on the future event horizon IR cut-off scenario.
However, to date, there is no observational evidence confirming the existence of torsion, making it difficult to specify its form.
For this reason, we will analyze three cases for the torsion scalar: (i) steady-state, (ii) constant, and (iii) time-dependent.

In case (i), where $\phi/H$ is constant, known as the steady-state torsion, the density parameter $\Omega_{X}$ becomes constant according to \eqref{effective density parameter_Hubble}, which causes $\omega_{X}$ to behave like $\omega_{m}$ via \eqref{equation of state_vacuum}.
This scenario must be ruled out.

\begin{figure}[h]
    \centering
    \includegraphics[width=0.94\linewidth]{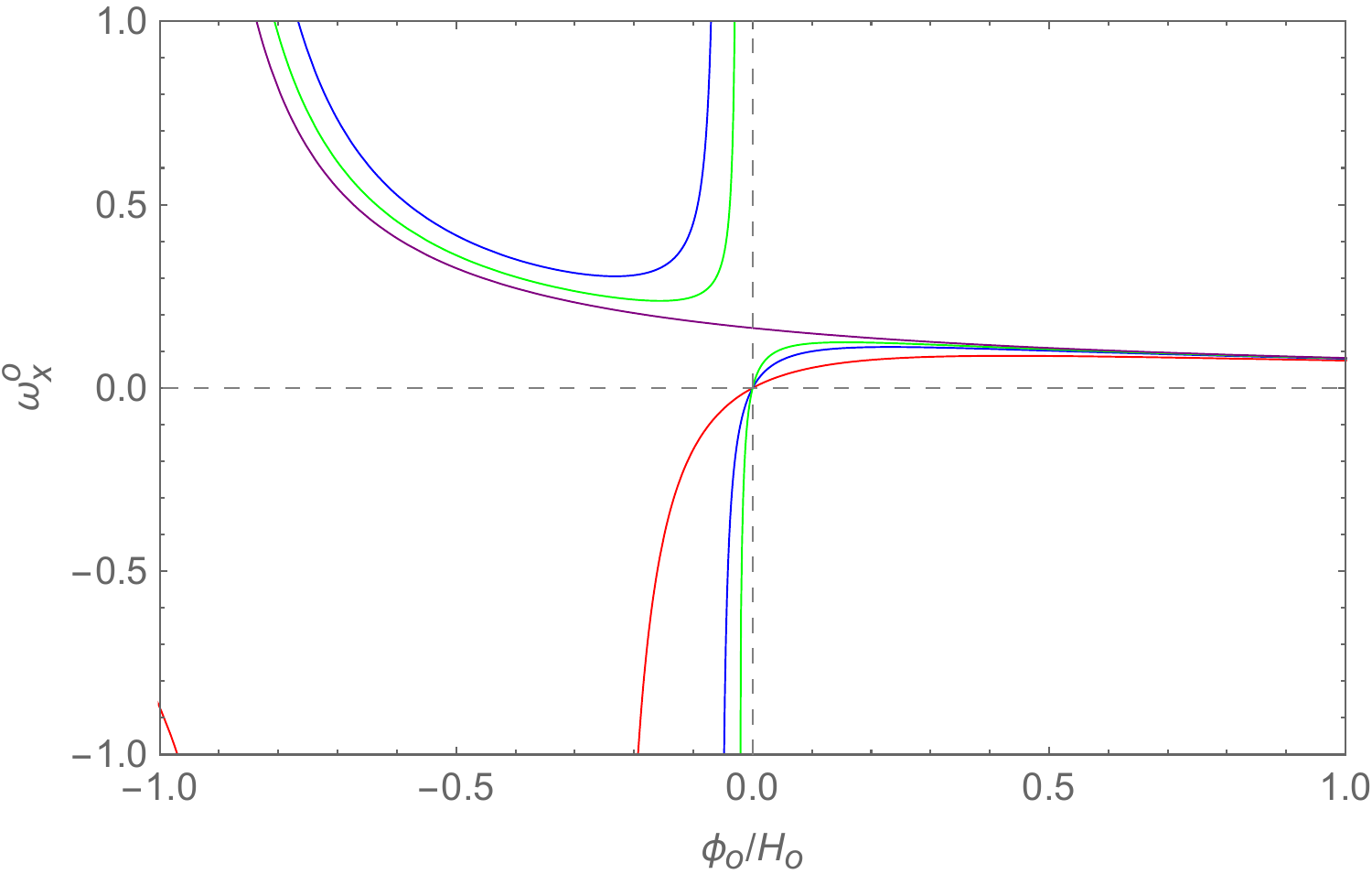}
    \caption{The current equation of state $\omega_{X}^{0}$ with the constant torsion scalar for various values of the free parameter $d = 0.5$ (red), $d = 0.886$ (blue), $d = 0.95$ (green), and $d = 1$ (purple).}
    \label{fig.1}
\end{figure}

In case (ii), where $\phi$ is constant, the equation of state \eqref{equation of state_Hubble} is simplified to 
\begin{equation} \label{constant case}
    \omega_{X} 
    = \omega_{m} + \frac{4}{3}\frac{\frac{\phi}{H}(1+q)}{\left(1+2\frac{\phi}{H}\right)^{2}-d^{2}}.
\end{equation}
We now evaluate the equation of state \eqref{constant case} at the present time, assuming dust with $\omega_{m} = 0$.
Instead of using an indirect value based on the $\Lambda$CDM model, we adopt the direct observational data for $q_{0} = -0.51$ from SH0ES (2022) \cite{18}, since 
the effect of the torsion scalar is reflected in the deceleration parameter through the Friedmann equation \eqref{deceleration parameter}.
As shown in Fig. \ref{fig.1}, some solutions support cosmic acceleration in weak torsion range $|\phi_{0}|/H_{0} < 1$, whereas others inhibit it.
Therefore, it is necessary to determine the exact value of $\phi_0$, but this remains challenging due to the current lack of observational data on torsion. In this regard, we can only assert the existence of solutions that allow cosmic acceleration.

In case (iii), where $\phi$ varies with time, many possible forms exist. 
Among these, a form of the torsion scalar that depends on the matter density appears to be a more realistic choice, since the spin of matter plays a role as the physical source of torsion.
According to the literature, the following form is suggested \cite{19}:
\begin{equation} \label{ansatz}
    \phi(t) 
    = -kH_{0}\left(\frac{H_{0}}{H(t)}\right)^{m}\left(\frac{\rho_{m}(t)}{3M_{p}^{2}H^{2}_{0}}\right)^{n}.
\end{equation}
Based on recent cosmological data, such as Type Ia supernovae and the Hubble parameter, the above parameters are determined to be $k=-0.30^{+0.51}_{-0.31}$, $m = -3.6^{+6.9}_{-7.6}$ and $n=-1.5^{+2.4}_{-2.7}$, where $k$ is a dimensionless parameter.
Motivated by the structure of \eqref{ansatz}, and assuming $m = -1$ and $n = 1$ to maintain some similarity with the literature \cite{19}, we propose the following ansatz within this framework:
\begin{equation} \label{ansatz 2}
    \phi(t)
    \simeq -kH(t)\left(\frac{\rho_{m}(t)}{3M_{p}^{2}H^{2}_{0}}\right).
\end{equation}
Here, the above ansatz does not represent a general solution.
As mentioned earlier, determining or inferring its specific form is difficult without direct observational data on torsion. Let us first focus on this particular ansatz to analyze whether a non-interacting holographic dark energy model that sets the Hubble radius as the IR cut-off is feasible.

To see this, we note that the evolution of the torsion scalar given in \eqref{ansatz 2} is governed by the continuity equation of matter \eqref{Non-interaction (matter)}, which results in
\begin{equation} \label{diff. non-constant scalar torsion}
    \dot{\phi} 
    \simeq - H\phi \left[ q+3+\left(1+3\omega_m\right)\left(1+2\frac{\phi}{H}\right) \right].
\end{equation}
From \eqref{equation of state_Hubble} with \eqref{diff. non-constant scalar torsion}, we arrive at
\begin{equation} \label{equation of state (non-zero scalar torsion)}
    \omega_{X} 
    \simeq \omega_{m}\left[1-4\frac{\frac{\phi}{H}\left(1+2\frac{\phi}{H}\right)}{\left(1+2\frac{\phi}{H}\right)^{2}-d^{2}}\right] 
   - \frac{4}{3}\frac{\frac{\phi}{H}\left(3+2\frac{\phi}{H}\right)}{\left(1+2\frac{\phi}{H}\right)^{2}-d^{2}}.
\end{equation}

\begin{figure}[h]
    \centering
    \includegraphics[width=0.94\linewidth]{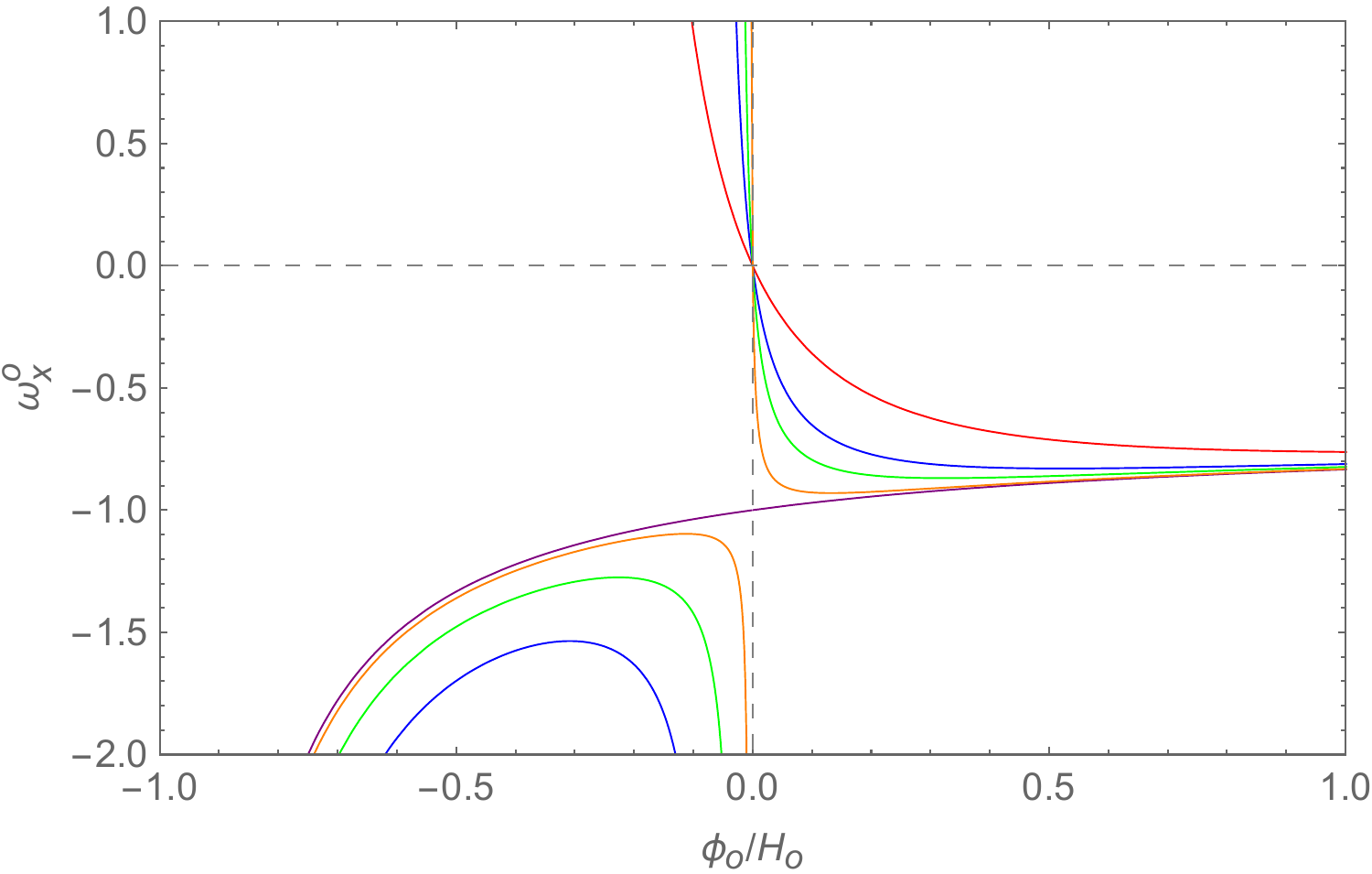}
    \caption{Possible solutions of $\omega_{X}^{0}$ with the time-dependent torsion scalar \eqref{ansatz 2} for various values of the free parameter $d = 0.5$ (red), $d = 0.886$ (blue), $d = 0.95$ (green), and $d = 1$ (purple).}
    \label{fig.2}
\end{figure}

\begin{figure}[h]
    \centering
    \includegraphics[width=0.94\linewidth]{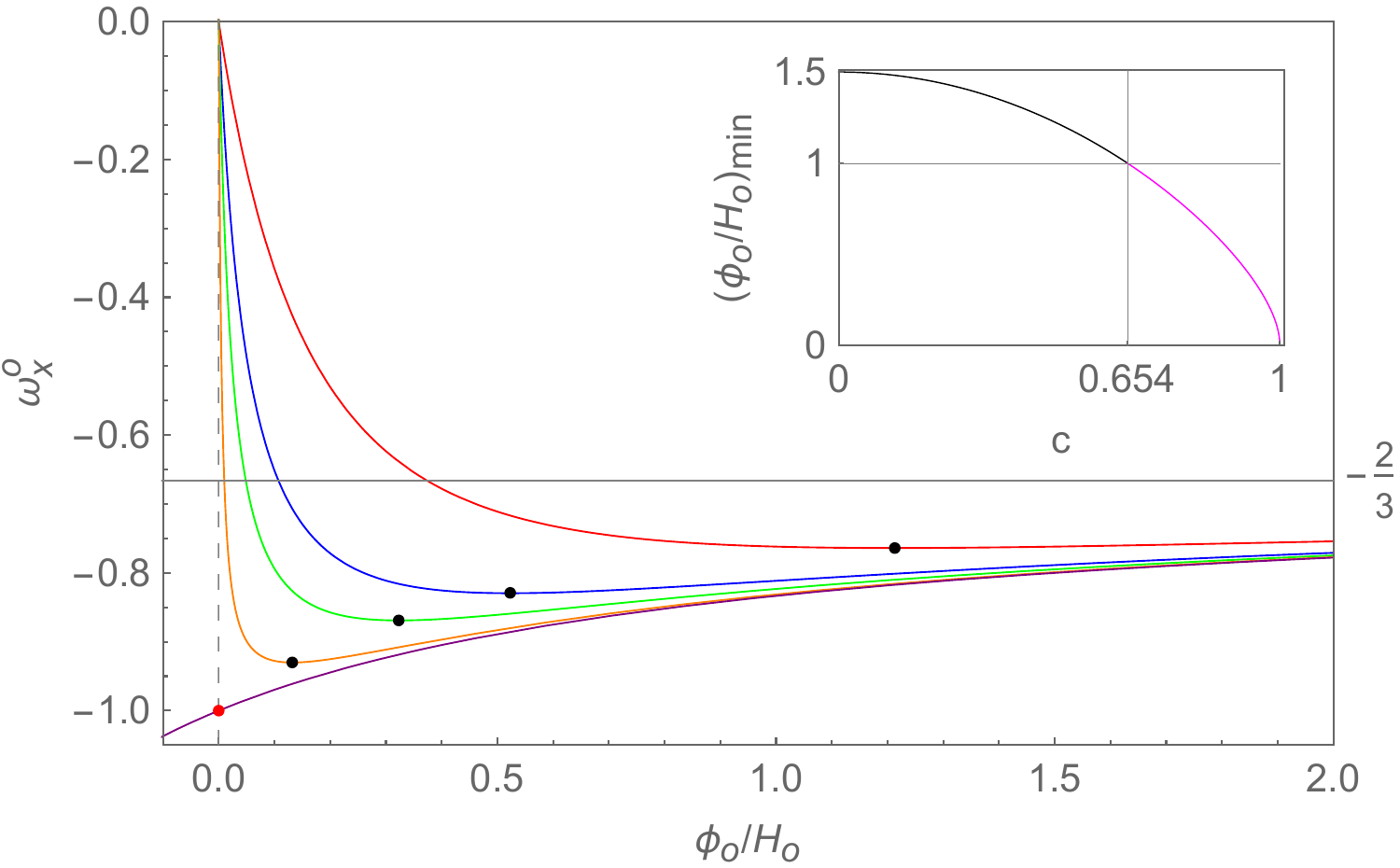}
    \caption{Close-up view of Fig. \ref{fig.2} for $\phi_{0}/H_{0} \geq 0$, highlighting the minima of $\omega_{X}^{0}$ (black dots) located at $(\phi_{0}/H_{0})_{min}$. For $d = 1$, the red dot represents $\omega_X^{0} = -1$ at $\phi_{0}/H_{0} = 0$, which is not a minimum. The inset illustrates the dependence of $(\phi_{0}/H_{0})_{min}$ on $d$, where the pink solid line indicates valid solutions within the weak torsion range for $0.654 < d < 1$.}
    \label{fig.3}
\end{figure}

To simplify the analysis, we set $\omega_{m} = 0$.
At the present time, given the negative ratio $\phi_{0}/H_{0}$ yields two types of solutions: one behaves like phantom dark energy with $\omega_{X}^{0} < -1$, while the other results in no acceleration, characterized by $\omega_{X}^{0} > 0$.
Therefore, it cannot be an appropriate candidate.
On the other hand, given the positive ratio, the equation of state lies within the range $-1 < \omega^{0}_{X} < 0$, as shown in Fig. \ref{fig.2}.
As a result, some solutions lead to cosmic acceleration.
Clearly, there exist minima $(\omega_{X}^{0})_{min}$ at $(\phi_{0}/H_{0})_{min}$ for various $d \neq 1$, as shown in Fig. \ref{fig.3}. 
We exclude the value $d = 1$, since it is unphysical in the torsion-free limit, and find
\begin{equation} \label{minima 1}
    \left(\frac{\phi_{0}}{H_{0}}\right)_{min} = \frac{1}{2}\left[1-d^{2}+\sqrt{\left(1-d^{2}\right)\left(4-d^{2}\right)}\right],
\end{equation} 
\begin{equation} \label{minima 2}
    (\omega_{X}^{0})_{min} = -\frac{1}{3}\left[2+\frac{1}{2-d^2+\sqrt{\left(1-d^{2}\right)\left(4-d^{2}\right)}}\right].
\end{equation}

\begin{table}[ht]
\caption{Minima of $\phi_{0}/H_{0}$ and $\omega^{0}_{X}$ for various values of $d$ in the weak torsion range.\label{table}}
{
\resizebox{\columnwidth}{!}{%
\begin{tabular}{@{}cccccc@{}}
\toprule
 & $d=0.654$ & $d=0.886$ & $d=0.95$ & $d=0.99$ & $d=1.0$ \\
\colrule
$(\phi_{0}/H_{0})_{\min}$ & $\textcolor{white}{+}1$ & $\textcolor{white}{+}0.523$ & $\textcolor{white}{+}0.323$ & $\textcolor{white}{+}0.132$ & $\textcolor{white}{+}0$ \\
$(\omega^{0}_{X})_{\min}$ & $-0.778$ & $-0.829$ & $-0.869$ & $-0.930$ & $-1$ \\
\botrule
\end{tabular}
}%
}
\end{table}

As shown in Table \ref{table}, the validity of the weak torsion is justified in a range of $0.654 < d < 1$ via \eqref{minima 1} and \eqref{minima 2}, and hence we find $(\omega_X^0)_{min} < \omega_X^0 < -2/3$ as $\phi_{0}/H_{0}$ goes from $(\phi_{0}/H_{0})_{min}$ to $\infty$.
If $(\phi_{0}/H_{0})_{min}$ is not too small, we obtain a range of $-1< (\omega^{0}_{X})_{min} <-0.778$  as $d$ changes from $1$ to $ 0.654$. 
As mentioned, focusing on the free parameter $d \approx 1$, valid solutions that are expected to match the observations should be found near the red dot in the Fig. \ref{fig.3}.

Therefore, even if torsion exists, its contribution becomes very small according to \eqref{minima 1}, and simultaneously $(\omega_{X}^{0})_{min}$ exhibits behavior that differs slightly from that of the cosmological constant.
In particular, the equation of state \eqref{equation of state (non-zero scalar torsion)} is distinct from the cosmological constant in that it is dynamic.
Although this analysis is only valid for the torsion scalar given in the ansatz \eqref{ansatz 2}, it shows the possibility that holographic dark energy model with torsion could set the Hubble radius as the IR cut-off point even in the absence of interaction between dark energy and dark matter.
This is noteworthy because it is not allowed within the framework of general relativity.

\section{Conclusions}
In the Friedmann cosmology with the torsion scalar, where there is no interaction between dark energy and dark matter, we have derived the equation of state for the holographic dark energy by setting the Hubble radius as the IR cut-off.
Here, we have analyzed three scenarios: the steady-state torsion, the constant torsion scalar, and the time-dependent torsion scalar linked to matter density containing spin.

The steady-state torsion behaves like matter and therefore fails to explain cosmic acceleration.
In contrast, the constant torsion scalar allows for both cosmic acceleration and non-acceleration.

In the regime dominated by the time-dependent torsion scalar resulting from the spin of matter, when it is non-negative, the equation of state is restricted to $-1 < \omega_{X}^{0} < 0$ in the weak torsion range. 
There exist minima of $\omega^{0}_{X}$, which lies within $-1 < (\omega_X^{0})_{min} < -0.778$ as the free parameter changes from $1$ to $ 0.654$.
The literature infers that the free parameter will be very close to unity.
Hence, the contribution of the torsional scalar in the expanding universe becomes very small, and $(\omega_X^{0})_{min}$ exhibits slightly different behavior from the cosmological constant.
Unlike the cosmological constant, this equation of state is dynamic.

These results are based on a special ansatz where the torsion scalar is proportional to matter density, due to the lack of observational data on torsion. 
Because torsion is a macroscopic representation of spin, it seems quite reasonable that the torsion scalar could be proportional to matter density. 
Nevertheless, this approach is noteworthy because the introduction of torsion suggests that the Hubble radius could be a candidate for the IR cut-off without assuming an interaction between dark energy and dark matter.
In particular, this torsional result shows a non-interaction limit which is absent in interaction models that set the IR cut-off to the Hubble radius within the framework of general relativity. 
As a result, it can provide an alternative means to avoid pathological problems associated with the future event horizon, including causality and circular logic issues.

Instead of the torsion used in this work, if we use a torsion whose relationship to the spin of matter is more explicit, it is expected to provide theoretical support for a much more feasible holographic dark energy model.

\appendix
\section{Continuity equation}
To avoid confusion caused by the addition of the torsion scalar, we briefly introduce the process of deriving the continuity equation from the Friedmann equations.
Using the Hubble parameter, we can rewrite the Friedmann equations \eqref{Friedmann equation 1} and \eqref{Friedmann equation 2} as
\begin{equation} \label{A.1}
    H^{2} + 4\phi^{2} + 4H\phi
    = \frac{1}{3M_{p}^{2}}\rho,
\end{equation}
and
\begin{equation} \label{A.2}
    \dot{H} + H^{2} + 2\dot{\phi} + 2H\phi
    = - \frac{1}{6M_{p}^{2}}\left(\rho+3p\right).
\end{equation}
Combining the two equations above, we obtain the following relation:
\begin{equation} \label{A.3}
    \dot{H} + 2\dot{\phi}
    = - \frac{1}{2M_{p}^{2}}\left(\rho+p\right) + 2H\phi\left(1+2\frac{\phi}{H}\right).
\end{equation}
Differentiating the first Friedmann equation \eqref{A.1} with respect to time gives
\begin{equation} \label{A.4}
    \frac{\dot{\rho}}{3M_{p}^{2}}
    = 2H\left(1+2\frac{\phi}{H}\right)\left(\dot{H}+2\dot{\phi}\right).
\end{equation}
Putting \eqref{A.3} into \eqref{A.4}, and using \eqref{A.1}, we obtain
\begin{equation} \label{A.5}
    \dot{\rho}
    = -3H\left(\rho+p\right)\left(1+2\frac{\phi}{H}\right) + 4\phi\rho,
\end{equation}
which can be written as
\begin{equation} \label{A.6}
    \dot{\rho} + 3H\left(\rho+p\right) + 2\phi\left(\rho+3p\right)
    = 0.
\end{equation}


\begin{acknowledgments}
This work was supported by the Daejin University Research Grants in 2025.
\end{acknowledgments}



\end{document}